\documentstyle[12pt,psfig]{article}
\begin{document}
\setlength{\headheight}{0in}
\setlength{\headsep}{0in}
\setlength{\topskip}{1ex}
\setlength{\textheight}{8.5in}
\setlength{\topmargin}{0.5cm}
\setlength{\baselineskip}{0.24in}
\catcode`@=11
\long\def\@caption#1[#2]#3{\par\addcontentsline{\csname
  ext@#1\endcsname}{#1}{\protect\numberline{\csname
  the#1\endcsname}{\ignorespaces #2}}\begingroup
    \small
    \@parboxrestore
    \@makecaption{\csname fnum@#1\endcsname}{\ignorespaces #3}\par
  \endgroup}
\catcode`@=12
\def\slashchar#1{\setbox0=\hbox{$#1$}           
   \dimen0=\wd0                                 
   \setbox1=\hbox{/} \dimen1=\wd1               
   \ifdim\dimen0>\dimen1                        
      \rlap{\hbox to \dimen0{\hfil/\hfil}}      
      #1                                        
   \else                                        
      \rlap{\hbox to \dimen1{\hfil$#1$\hfil}}   
      /                                         
   \fi}                                         %
\newcommand{\newc}{\newcommand}
\def\be{\begin{equation}}
\def\ee{\end{equation}}
\def\bea{\begin{eqnarray}}
\def\eea{\end{eqnarray}}
\def\simlt{\stackrel{<}{{}_\sim}}
\def\simgt{\stackrel{>}{{}_\sim}}
\begin{titlepage}
\begin{flushright}
{\setlength{\baselineskip}{0.18in}
{\normalsize
IC/99/1\\
hep-ph/9901389\\
January 1999\\
}}
\end{flushright}
\vskip 2cm
\begin{center}

{\Large\bf 
Effects of the supersymmetric phases on the neutral Higgs sector}

\vskip 1cm

{\large
D. A. Demir \footnote{e-mail: ddemir@ictp.trieste.it} \\}

\vskip 0.5cm
{\setlength{\baselineskip}{0.18in}
{\normalsize\it The Abdus Salam International Center for Theoretical Physics, I-34100,
Trieste, Italy \\}}

\end{center}
\vskip .5cm
\begin{abstract}
By using the effective potential approximation and taking into account the dominant top quark and scalar top quark loops,
radiative corrections to MSSM Higgs potential are computed in the presence of the supersymmetric CP--violating phases.
It is found that, the lightest Higgs scalar remains essentially CP--even as in the CP--invariant theory whereas the other
two scalars are heavy and do not have definite CP properties. The supersymmetric CP--violating phases are shown to modify
significantly the decay rates of the scalars to fermion pairs.

\end{abstract}
\end{titlepage}
\setcounter{footnote}{0}
\setcounter{page}{1}
\setcounter{section}{0}
\setcounter{subsection}{0}
\setcounter{subsubsection}{0}

\section{Introduction}
The Minimal Supersymmetric Standard Model (MSSM) consists of various soft supersymmetry
breaking parameters as well as the Higgsino mass parameter, $\mu$, coming from the  
superpotential. In general, there is no a priori reason for taking all these parameters real,
and thus, Yukawa couplings, gaugino masses, trilinear Higgs--sfermion couplings, 
$A_{f}$, Higgs bilinear coupling, $m_{3}^{2}\equiv \mu B$, and $\mu$-- parameter itself can all
be complex. On the other hand, the MSSM Lagrangian has two global symmetries $U(1)_{PQ}$ 
(Peccei-Quinn symmetry) and $U(1)_{R-PQ}$ (an R symmetry) under which all fields and 
parameters are charged. The selection rules for these symmetries limit the combinations of 
dimensionful parameters that can appear in a physical quantity so that one has, in fact, only 
three of these phases physical \cite{phase1,phase2}. Without loss of generality, these three 
physical phases can be identified with (1) the phase in the
Cabibbo--Kobayashi--Maskawa matrix, $\delta_{CKM}$, (2) $\varphi_{\mu}\equiv
{\cal{A}}rg(\mu)$, and (3) $\varphi_{A_{f}}\equiv {\cal{A}}rg(A_{f})$. Thus any physical quantity
${\cal{F}}$ has an explicit dependence on these phases: ${\cal{F}}={\cal{F}}(\delta_{CKM}, 
\varphi_{\mu}, \varphi_{A_{f}})$.

Despite their presence in the Lagrangian, the phenomenological relevance of
these CP--violating phases has often been questioned due to the smallness of the
neutron and electron electric dipole moments \cite{phase1,edm} which require them to be
at most ${\cal{O}}(10^{-3})$. However, recent studies have shown that it is possible to
suppress neutron and electron dipole moments without requiring these CP-violating phases to be
small by allowing the existence of either non-universal soft breaking parameters at the
unification scale \cite{non-uni} or some kind of cancellation among various
supersymmetric contributions \cite{cancel}, or heavy enough sfermions for the first two
generations \cite{heavy12}. In fact, following the last scenario, it was recently shown
that the CP--violation in B-- and K-- systems can be saturated with $\varphi_{\mu}$ and
$\varphi_{A_{f}}$ only \cite{biz}. Apart from electric dipole moments and weak decays, 
these phases play a crucial role in the creation of the baryon asymmetry of the universe 
at the electroweak phase transition \cite{phase}.

In this work, assuming that the electric dipole moments are suppressed by one of the  
methods metioned above, we take supersymmetric phases unconstrained, 
and investigate their effects on the Higgs sector of the MSSM. At the tree-level
the Higgs sector of the MSSM conserves CP due to the fact that the superpotential 
is holomorphic in superfields entailing the absence of flavour changing 
neutral currents and scalar--pseudoscalar mixings. When the supersymmetric phases
$\varphi_{\mu}$ and $\varphi_{A_{f}}$ vanish Higgs sector conserves CP at any loop 
order. In fact, the CP--conserving Higgs sector has been analyzed 
by several authors with the main purpose of evaluating the mass of the lightest Higgs 
boson which has the tree-level upper bound of $M_{Z}$. It has been found that
radiative corrections, dominated by top and stop loops, elevate the tree-level 
bound significantly \cite{radiative1}. These one-loop results \cite{radiative1}
have been improved by utilizing complete one-loop on-shell renormalization 
\cite{on-shell}, renormalization group methods \cite{renorm}, diagrammatic methods 
with leading order QCD corrections \cite{diag}, and two-loop on-shell renormalization 
\cite{hollik}. However, when the supersymmetric phases are non-vanishing, as the 
recent studies have shown \cite{pilaftsis}, the Higgs sector become CP-- violating 
through the radiative corrections. As in the CP--conserving case, the radiative  
corrections will be dominated by the top and stop loops. Below we investigate 
effects of the supersymmetric phases on the Higgs masses,  scalar-pseudoscalar mixings, and 
decay properties of scalars to fermion pairs. In doing this, we shall calculate one-loop
radiative corrections coming from top and stop loops in the effective potential 
approximation. 

This work is organized as follows. In Sec. 2 we compute the one-loop effective potential 
using top quark and stop contributions together with the specification of the particle 
spectrum and mixings. In Sec. 3 we discuss, as an example, the decay properties of 
the Higgs scalars to fermion pairs. In Sec. 4 we conclude the work.

\section{Effective potential}
The Higgs sector of the MSSM consists of two SU(2) doublets $H_{1}$, $H_{2}$ with opposite 
hypercharges $Y_{1}=-1$, $Y_{2}=+1$, and non-vanishing vacuum expectation values $v_{1}$, 
$v_{2}$. Allowing a finite alignment, $\theta$, between the two Higgs doublets, 
we adopt the following decomposition:
\bea
H_{1}&=&\left(\begin{array}{c c} H_{1}^{0}\\
H_{1}^{-}\end{array}\right)=\frac{1}{\sqrt{2}}\left(\begin{array}{c c}
v_{1}+\phi_{1}+i\varphi_{1}\\ H_{1}^{-}\end{array}\right)\nonumber\\
H_{2}&=&\left(\begin{array}{c c} H_{2}^{+}\\
H_{2}^{0}\end{array}\right)=\frac{e^{i\theta}}{\sqrt{2}}\left(\begin{array}{c c}
H_{2}^{+}\\ v_{2}+\phi_{2}+i\varphi_{2}\end{array}\right)\; . 
\eea
At the tree level the Higgs sector is described by the scalar potential
\bea
V_{0}(H_{1}, H_{2})&=&m_{1}^{2}|H_{1}|^2+m_{2}^{2}|H_{2}|^2+
  (m_{3}^{2}H_{1}\cdot H_{2}+H.c.)\nonumber\\
&+&\frac{\lambda_{1}}{2}|H_{1}|^{4}+\frac{\lambda_{2}}{2}|H_{2}|^{4}+
\lambda_{12} |H_{1}|^2 |H_{2}|^2+\tilde{\lambda}_{12}|H_{1}\cdot H_{2}|^{2}   
\eea
with the parameters
\bea
m_{1}^{2}&=&m_{\tilde{H}_{1}}^{2}+|\mu|^{2}\; , \; m_{2}^{2}=m_{\tilde{H}_{2}}^{2}+|\mu|^{2}\; , \;
m_{3}^{2}=|\mu B|\; , \; \lambda_{1}=\lambda_{2}=(g_{2}^{2}+g_{1}^{2})/4\nonumber\\
\lambda_{12}&=&(g_{2}^{2}-g_{1}^{2})/4\; , \; \tilde{\lambda}_{12}=-g_{2}^{2}/2
\eea
where $m_{\tilde{H}_{1,2}}^{2}$ and $B$ are the soft supersymmetry breaking parameters. 
As is seen from (3), the tree level potential is described by real parameters; thus, the 
alignment between the two doublets can, in fact, be rotated away. Since, in the minimum,  
the potential is to have vanishing gradients in all directions, in particular, $\partial V_{0}/\partial
\varphi_{1,2} = m_{3}^{2} \sin \theta = 0$, one automatically gets $\theta=0$. Those terms of the tree
level potential (2) quadratic in the components of the Higgs doublets (1) 
give the mass-squared matrix of neutral scalars the diagonalization of which yields the 
CP=-1 boson $A^{0}=\cos\beta\varphi_{1}-\sin\beta\varphi_{2}$ with mass $M_{A^{0}}^{2}=
-m_{3}^{2}/\sin\beta\cos\beta$, and two CP=+1 bosons which are linear combinations of $\phi_{1}$ 
and $\phi_{2}$ with a mixing angle $\alpha$. The mixing angle $\alpha$ and the masses of the CP even
scalars $h$ and  $H$ are given by \cite{hunter}
\bea
\tan 2 \alpha &=& \frac{M_{A^{0}}^{2}+M_{Z}^{2}}{M_{A^{0}}^{2}-M_{Z}^{2}}\tan 2\beta\\
M_{h (H)}^{2} &=& \frac{1}{2}( M_{A^{0}}^{2}+M_{Z}^{2}-(+)\sqrt{(M_{A^{0}}^{2}+M_{Z}^{2})^{2}-4 M_{A^{0}}^{2}M_{Z}^{2}\cos^{2}
2\beta})
\eea
where $\tan\beta\equiv v_{2}/v_{1}$, and $M_{Z}^{2}=(g_{2}^{2}+g_{1}^{2})(v_{1}^{2}+v_{2}^{2})/4$ in our convention.
It is readily seen that for $\tan\beta \simgt 2$ one has $\beta\sim \pi/2$, $M_{h}\sim M_{Z}$ and $M_{H}\sim M_{A}$. However, it
is known that radiative corrections elevate $M_{h}$ (bounded by $M_{Z}$ at the tree level) significantly \cite{radiative1}
without modifying the mass degeneracy between $H$ and $A$ when the theory conserves CP. When, however, the CP--violating MSSM
phases are switched on, the degeneracy between $H$ and $A$ can be lifted considerably as discussed in \cite{pilaftsis}.

We now start computing the radiative corrections to the tree potential (2) in the presence of the CP--violating 
MSSM phases $\varphi_{\mu, A_{f}}$. Our main concern will be the investigation of the masses and mixings of the scalars 
as a function of the CP--violating angles. These mixings could be CP--conserving (like $h-H$ mixing in the CP--respecting limit)
as well as CP--violating as we will discuss below. To evaluate the radiative corrections we follow the effective potential 
approximation where the tree-level potential (2) is added to the one-loop contributions having the famous 
Coleman-Weinberg \cite{radiative1,sher} form
\bea
\Delta V = \frac{1}{64 \pi^{2}} Str
{\cal{M}}^{4}(H_{1},H_{2})(\log{\frac{{\cal{M}}^{2}(H_{1},H_{2})}{Q^{2}}}-\frac{3}{2})
\eea
where $Str\equiv \sum_{J} (-1)^{2 J +1} (2 J + 1)$ is the usual supertrace, and
${\cal{M}}(H_{1},H_{2})$ is the Higgs field dependent mass matrix of particles. $\Delta V$
depends on the renormalization scale $Q$ which is presumably around the weak scale. 
In evaluating $\Delta V$ one includes the contributions of vector bosons, Higg bosons and
fermions as well as their supersymmetric partners gauginos, Higgsinos and sfermions. Among all
these particles top quarks and scalar top quarks give the dominant contributions
\cite{radiative1}. However, for very large $\tan \beta$ values
bottom-sbottom and $\tau$-stau systems can become important. Besides these, since the dependence
of $\Delta V$ on the CP-violating phases $\varphi_{\mu, A_{f}}$
originates from only the Higgsino (through $\mu$ dependence) and sfermion (through $\mu$ and $A_{f}$
dependence) mass matrices, these two particle species attain a seperate importance. However, if one wishes
to include the Higgsino contribution, all particle species must be included since then precision of the
computation rises to the level of gauge couplings. In the following we neglect the contributions of
gauge couplings and restrict ourselves to moderate values of $\tan\beta$ so that, to a good approximation, 
the dominant terms in $\Delta V$ are given by top-- stop system. This approximation is convenient in that it picks
up the phase-sensitive dominant contributions to $\Delta V$.  

In $(\tilde{t}_{L}, \tilde{t}_{R})$ basis stop mass-squared matrix, neglecting the D-term
contributions, takes the form  
\bea
M_{\tilde{t}}=\left(\begin{array}{c c} M_{\tilde{L}}^{2}+h_{t}^{2}{|H_{2}^{0}|}^{2} &
h_{t}(A_{t}H^{0}_{2}-\mu^{*}{H_{1}^{0}}^{*})\\
h_{t}(A_{t}^{*}{H^{0}_{2}}^{*}-\mu H_{1}^{0}) & M_{\tilde{R}}^{2}+h_{t}^{2}{|H_{2}^{0}|}^{2}
\end{array}\right)
\eea    
where $\mu$ and $A_{t}$ are complex, $M_{\tilde{L},\tilde{R}}^{2}$ are the soft 
mass-squareds of left-- and right--handed stops, and $h_{t}$ is the top Yukawa coupling.
Denoting the eigenvalues of $M_{\tilde{t}}$ by $m_{\tilde{t}_{1,2}}^{2}$ and using 
$m_{t}^{2}= h_{t}^{2} {|H_{2}^{0}|}^{2}$ the one-loop effective potential takes the form 
\bea
V=V_{0}+\frac{6}{64 \pi^{2}}\left( \sum_{a=\tilde{t}_{1},\tilde{t}_{2}} m_{a}^{4} ( \log{
\frac{m_{a}^{2}}{Q^{2}}}-\frac{3}{2})- 2 m_{t}^{4}( \log{ 
\frac{m_{t}^{2}}{Q^{2}}}-\frac{3}{2})\right) \; .
\eea
We require this effective potential to be minimized at $(v_{1}, v_{2}, \theta)$ at which 
it has to have vanishing gradients in all directions and the masses of the Higgs scalars 
must be real positive. Gradients of the potential with respect to 
the charged components of the Higgs doublets automatically vanish as there is no charge 
breaking effects in the vacuum. On the other hand, extremization of $V$ with respect to neutral 
components of the Higgs doublets yield
\bea
&&v_{1}(2 m_{1}^{2} +\lambda_{1} v_{1}^{2} + (\lambda_{12}+\tilde{\lambda}_{12})v_{2}^{2}) + 2
m_{3}^{2} v_{2} \cos \theta + 2 \left (\frac{\partial \Delta V}{\partial \phi_{1}}\right)_{0}
=0\\ 
&&v_{2}(2 m_{2}^{2} +\lambda_{2} v_{2}^{2} + (\lambda_{12}+\tilde{\lambda}_{12})v_{1}^{2}) + 2
m_{3}^{2} v_{1} \cos \theta + 2 \left (\frac{\partial \Delta V}{\partial \phi_{2}}\right)_{0}    
=0\\
&&m_{3}^{2} v_{2} \sin \theta - \left (\frac{\partial \Delta V}{\partial \varphi_{1}}\right)_{0}
=0\\
&&m_{3}^{2} v_{1} \sin \theta - \left (\frac{\partial \Delta V}{\partial \varphi_{2}}\right)_{0}
=0
\eea
where the subscript $"0"$ implies the substitution $\phi_{1}=\phi_{2}=\varphi_{1}=\varphi_{2}=0$ 
in the corresponding quantity. Equations (9) and (10) come by no surprise as they are the
counterparts of the ones occurring in the CP-conserving case. However, with $\left (\partial \Delta V/\partial
\varphi_{1,2}\right)_{0}\neq 0$, equations (11) and (12) now imply a non-trivial solution for $\sin\theta$
unlike the CP-conserving case where these gradients indentically vanish and one automatically obtains a
vanishing $\theta$. After some algebra one can show that 
\bea
\left (\frac{\partial \Delta V}{\partial \varphi_{1}}\right)_{0} = \tan \beta \left (\frac{\partial
\Delta V}{\partial \varphi_{2}}\right)_{0}\; .
\eea
Therefore, equations (11) and (12) imply one and the same solution for $\theta$
\bea
m_{3}^{2} \sin \theta =\frac{1}{2}\beta_{h_{t}} |\mu| |A_{t}| \sin \gamma
f(m_{\tilde{t}_{1}}^{2}, m_{\tilde{t}_{2}}^{2})
\eea
where $\beta_{h_{t}}=(3 h_{t}^{2})/16 \pi^{2}$, $\gamma = \varphi_{\mu}+\varphi_{A_{t}}$, and 
\bea
f(x,y)=-2+\log{\frac{xy}{Q^{4}}}+\frac{ y + x } { y - x} \log{\frac{y}{x}}
\eea
is a scale-dependent one-loop function. Actually,
if one uses the decomposition of the Higgs doublets in (1), in all one-loop formulae $\gamma$
gets replaced by $\gamma + \theta$. However, since $\theta$ itself is a loop-induced quantity,
its appearence together with $\gamma$ is a two-- and higher-- loop effect which we neglect.
Thus, when computing $\Delta V$ we drop $\theta$ from (1) knowing that it is induced through
(14). It is readily seen from (14) that unless $\gamma$ vanishes $\theta$ remains finite.
Furthermore, due to the form of the stop mass-squared matrix, all one-loop quantities turn out
to depend on the combination $\gamma = \varphi_{\mu}+\varphi_{A_{t}}$. Of course, had we included
the Higgsino contributions there would be terms that depend solely on $\varphi_{\mu}$ destructing
this kind of relation. However, they would be subleading compared to top and stop contributions 
discussed here.
 
In (14) and all formulae below $m_{\tilde{t}_{1,2}}^{2}$ denote stop mass-squared eigenvalues 
evaluated at the minimum of the potential: 
\bea
m_{\tilde{t}_{1,2}}^{2}=\frac{1}{2}\left ( M_{\tilde{L}}^{2}+M_{\tilde{R}}^{2}+ 2
m_{t}^{2} \mp \Delta_{\tilde{t}}^{2}\right)
\eea
where 
\bea
\Delta_{\tilde{t}}^{2}=\sqrt{(M_{\tilde{L}}^{2}-M_{\tilde{R}}^{2})^{2}+ 4 m_{t}^{2} (
|A_{t}|^{2}+
|\mu|^{2} \cot^{2} \beta-2 |\mu| |A_{t}| \cot \beta \cos \gamma)}\; .
\eea
 
As usual, construction of the mass-squared matrix of the Higgs scalars proceeds through the
evaluation  of
\bea
M^{2}=\left(\frac{\partial^{2} V} {\partial \chi_{i} \partial \chi_{j}}\right)_{0}\,,
\mbox{where}\; 
\chi_{i} \in {\cal{B}}=\{\phi_{1}, \phi_{2}, \varphi_{1}, \varphi_{2}\} \; .
\eea
Solving $m_{1,2}^{2}$ from the stationarity conditions (9) and (10), and replacing them in $M^{2}$ 
one observes that, in the basis ${\cal{B}}$, the vector $\{0, 0, -\cos\beta, \sin\beta \}$
corresponds to the Goldstone mode $G^{0}$ eaten by $Z$ boson to acquire its mass. Then in the
reduced basis ${\cal{B}}'=\{\phi_{1}, \phi_{2}, \sin\beta \varphi_{1}+\cos\beta \varphi_{2}\}$ the
mass-squared matrix of the Higgs scalars becomes 
\bea
M^{2}=\left(\begin{array}{c c c} M_{Z}^{2} c_{\beta}^{2} + \tilde{M}_{A}^{2} s_{\beta}^{2}+
\Delta_{11} &
-(M_{Z}^{2}+\tilde{M}_{A}^{2})s_{\beta}c_{\beta}+\Delta_{12} & r \Delta \\
-(M_{Z}^{2}+\tilde{M}_{A}^{2})s_{\beta}c_{\beta}+\Delta_{12} & M_{Z}^{2} s_{\beta}^{2} +
\tilde{M}_{A}^{2}
c_{\beta}^{2} + \Delta_{22} & s \Delta \\
r \Delta & s \Delta & \tilde{M}_{A}^{2}+\Delta
\end{array}\right)
\eea 
where $c_{\beta}=\cos\beta$ and $s_{\beta}=\sin \beta$. The scalar mass-squared matrix involves 
various parameters whose explicit expressions we list below. The adimensional parameters $r$ and
$s$ are given by 
\bea 
r&=&-\frac{\sin\beta}{\sin \gamma} \frac{|A_{t}|\cos\gamma-|\mu|\cot\beta}{|A_{t}|}\\
s&=&\frac{\sin\beta}{\sin\gamma}\frac{1}{|\mu||A_{t}|g(m_{\tilde{t}_{1}}^{2},
m_{\tilde{t}_{2}}^{2})}( |A_{t}|(|A_{t}|-|\mu|\cot\beta \cos\gamma)g(m_{\tilde{t}_{1}}^{2}, 
m_{\tilde{t}_{2}}^{2})\nonumber\\&-&(m_{\tilde{t}_{2}}^{2}-m_{\tilde{t}_{1}}^{2})\log{
\frac{m_{\tilde{t}_{2}}^{2}}{m_{\tilde{t}_{1}}^{2}}}) \;. 
\eea
While $r$ originates mainly from the stop left-right mixings, $s$ has an additional term 
depending on the the stop mass splitting $\log{\frac{m_{\tilde{t}_{2}}^{2}}{m_{\tilde{t}_{1}}^{2}}}$. 

The parameters of mass dimension can be expressed as follows
\bea
\Delta &=& -2 \beta_{h_{t}}\frac{\sin^{2}\gamma}{\sin^{2} \beta}\frac{|\mu|^{2} |A_{t}|^{2}
m_{t}^{2}}{(m_{\tilde{t}_{2}}^{2}-m_{\tilde{t}_{1}}^{2})^{2}} g(m_{\tilde{t}_{1}}^{2},
m_{\tilde{t}_{2}}^{2})\\
\Delta_{11} &=& -2\beta_{h_{t}}\frac{|\mu|^{2}  
m_{t}^{2} (|A_{t}|\cos\gamma - |\mu|\cot\beta)^{2}}
{(m_{\tilde{t}_{2}}^{2}-m_{\tilde{t}_{1}}^{2})^{2}} g(m_{\tilde{t}_{1}}^{2},
m_{\tilde{t}_{2}}^{2})\\
\Delta_{12} &=& -2\beta_{h_{t}} |\mu|m_{t}^{2}\{\frac{|A_{t}|\cos\gamma -|\mu|\cot\beta}
{(m_{\tilde{t}_{2}}^{2}-m_{\tilde{t}_{1}}^{2})} \log{  
\frac{m_{\tilde{t}_{2}}^{2}}{m_{\tilde{t}_{1}}^{2}}}\nonumber\\&-&
\frac{|A_{t}|[            
(|A_{t}|\cos\gamma - |\mu|\cot\beta)^{2}+|A_{t}|(|A_{t}|-|\mu|\cot\beta)\sin^{2} \gamma ]}
{(m_{\tilde{t}_{2}}^{2}-m_{\tilde{t}_{1}}^{2})^{2}}\nonumber\\&\times& g(m_{\tilde{t}_{1}}^{2},
m_{\tilde{t}_{2}}^{2})\}\\
\Delta_{22} &=& 2\beta_{h_{t}} m_{t}^{2}\{ \log { \frac{m_{\tilde{t}_{2}}^{2}
m_{\tilde{t}_{1}}^{2}}{m_{t}^{4}}}+\frac{2|A_{t}|(|A_{t}|-|\mu|\cot\beta\cos\gamma)}
{(m_{\tilde{t}_{2}}^{2}-m_{\tilde{t}_{1}}^{2})} \log{
\frac{m_{\tilde{t}_{2}}^{2}}{m_{\tilde{t}_{1}}^{2}}}\nonumber\\
&-&\frac{|A_{t}|^{2}(|A_{t}| - |\mu|\cot\beta\cos\gamma)^{2}}
{(m_{\tilde{t}_{2}}^{2}-m_{\tilde{t}_{1}}^{2})^{2}} g(m_{\tilde{t}_{1}}^{2},
m_{\tilde{t}_{2}}^{2})\}
\eea
where the function $g(m_{\tilde{t}_{1}}^{2}, m_{\tilde{t}_{2}}^{2})$ in these expressions reads
\bea
g(x,y)=f(x,y)-\log{\frac{xy}{Q^{4}}}.
\eea 
Therefore, unlike $f(m_{\tilde{t}_{1}}^{2}, m_{\tilde{t}_{2}}^{2})$, $g(m_{\tilde{t}_{1}}^{2},
m_{\tilde{t}_{2}}^{2})$ does not have an explicit dependence on the renormalization scale $Q$. In fact, the 
adimensional parameters $r$ and $s$ as well as the mass parameters (21)-(24) have no explicit
dependence on $Q$. On the other hand, the remaining mass parameter, $\tilde{M}_{A}^{2}$, 
in the scalar mass-squared matrix (19) is an explicit function of $Q$: 
\bea
\tilde{M}_{A}^{2}=\frac{m_{3}^{2}}{\sin\beta\cos\beta}\frac{\sin(\theta-\gamma)}{\sin\gamma}\; .
\eea
It is the $\theta$ (13) dependence of $\tilde{M}_{A}^{2}$ that makes it $Q$-dependent. However, the
explicit $Q$ dependence of $\theta$ should cancel with the implicit $Q$ dependence of $\tan\beta$ and
$m_{3}^{2}$ to make $\tilde{M}_{A}^{2}$ scale-independent \cite{radiative1}.  
 
According to the decomposition of Higgs doublets in (1), $\phi_{1}$ and $\phi_{2}$ are of CP=+1
whereas $\sin\beta \varphi_{1}+\cos\beta\varphi_{2}$ is of CP=-1. As suggested by the form of the 
scalar mass-squared matrix (19) there are mainly two kinds of mixings: (1) mixing of the scalars 
with different CP induced by $M^{2}_{13}=r \Delta$ and $M^{2}_{23}=s \Delta$, and (2) mixing of the 
CP=+1 scalars through $M_{12}^{2}=-(M_{Z}^{2}+\tilde{M}_{A}^{2})s_{\beta}c_{\beta}+\Delta_{12}$.  
While the former are induced purely by the non-vanishing supersymmetric phases the latter exists 
in the CP--respecting limit too. In the CP--conserving limit, that is, $\sin \gamma \rightarrow 0$, 
one obtains $r\Delta\rightarrow 0$, $s\Delta\rightarrow 0$ and $\Delta\rightarrow 0$. In this 
case CP=+1 and CP=-1 sectors in (19) decouple and reproduce the particle spectrum of the CP-- conserving
limit in which $\tilde{M}_{A}$ becomes the radiatively corrected pseudoscalar mass, and
$\Delta_{11,12,22}$ become the usual one-loop contributions \cite{radiative1} to the CP=+1
scalar mass-squared matrix. For a proper interpretation of results of the numerical analysis below
it is convenient to know the relative strengths of the CP--violating and conserving mixings. 
For large $\tan\beta$ equations (20) and (21) go over to
\bea
r\sim -\cot\gamma\;, \;\;\;\; s\sim \frac{1}{\sin\gamma}(\frac{|A_{t}|}{|\mu|}+\frac{4
(m_{t}^{2}+M_{Q}^{2})}{|\mu||A_{t}|})
\eea
where we assumed $M_{\tilde{L}}\sim M_{\tilde{R}}\equiv M_{Q}^{2}>> m_{t}|A_{t}|$ and $|A_{t}|>> |\mu|\cot\beta$ in derivation. 
Equation (28), thus,  implies that $|r|/|s| < <1$. This follows mainly from the dependence of the stop mass-matrix (7) on $H_{2}^{0}$ 
which causes not only $|s \Delta|$ but also $|\Delta_{12}|$ and $|\Delta_{22}|$ to be larger than $|\Delta_{11}|$ and $|r \Delta|$
through stop and stop-top splittings. An immediate consequence of (28) is that one eigenstate of the scalar mass-squared 
matrix (19) will be of mainly CP=+1. Thus one expects that among the three mass eigenstate scalars one will continue 
to have CP=+1 with a small CP=-1 component while the other two can mix significantly depending on the relative stregths 
of the other one-loop corrections. This observation can be justified numerically by analyzing the relative strengths 
of CP--violating and CP--conserving mixings as classified above. Fig.1 shows the variation of
$M^{2}_{13}/|M^{2}_{12}|$ (solid curve) and $M^{2}_{23}/|M^{2}_{12}|$ (dashed curve) with $\tan\beta$
for $M_{\tilde{L}}=M_{\tilde{R}}=|A_{t}|=10\cdot M_{Z}$, $|\mu|=2.5\cdot M_{Z}$ and $\tilde{M}_{A}=2\cdot
M_{Z}$ with $\gamma=\pi/4$. As the figure suggests larger the $\tan\beta$ bigger the CP--violating mixings
compared to the mixing between the CP=+1 components. The increase of these ratios with $\tan\beta$ can be understood as follows: (1)
$\beta\rightarrow \pi/2$ as $\tan\beta$ grows to higher values, and thus,  $|-(M_{Z}^{2}+\tilde{M}_{A}^{2})s_{\beta}c_{\beta}|$
decreases gradually, (2) $\Delta_{12}$ is positive for these parameters (and even for higher values of $|\mu|$ due to $\cot\beta$
suppression) and grows with $\tan\beta$ due to $\log{m_{\tilde{t}_{2}}^{2}/m_{\tilde{t}_{1}}^{2}}$ so that $M_{12}^{2}$ decreases with
increasing $\tan\beta$, and the ratios increase gradually since $M^{2}_{13}$ and $M^{2}_{23}$ decrease with $\tan\beta$ more slowly. 

\begin{figure}[hbt]
\centerline{
\psfig{figure=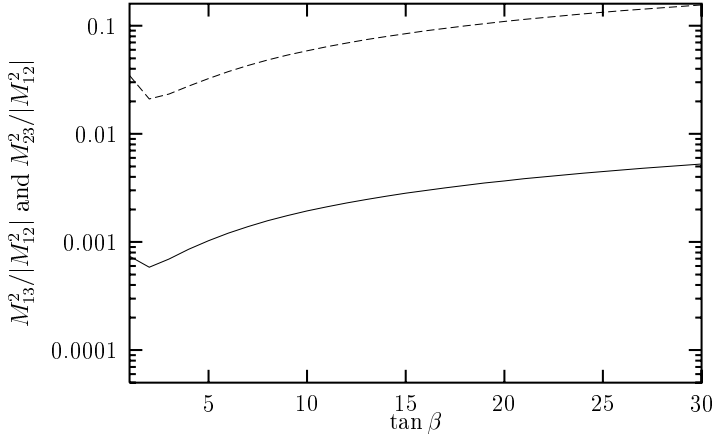,height=10cm,bbllx=1.0cm,bblly=9.cm,bburx=24.cm,bbury=22.cm}}
\caption{Variation of $M^{2}_{13}/|M^{2}_{12}|$ (solid curve) and $M^{2}_{23}/|M^{2}_{12}|$ (dashed curve)
with $\tan\beta$ for $M_{\tilde{L}}=M_{\tilde{R}}=|A_{t}|=10\cdot M_{Z}$, $|\mu|=2.5\cdot M_{Z}$ and
$\tilde{M}_{A}=2\cdot M_{Z}$ with $\gamma=\pi/4$. The CP-- violating mixings become important for large 
$\tan\beta$.}
\end{figure}

In Fig. 1 we plot ratios of the CP--violating mixings to CP--conserving ones; however, if one plots 
CP--violating mixings directly, for example in units of $(10\cdot M_{Z})^{2}$ for $|\mu|=10\cdot M_{Z}$, 
$\gamma=\pi/2$ and $\tilde{M}_{A}=5\cdot M_{Z}$, in absolute magnitude, $M^{2}_{13}$ ($M^{2}_{23}$) starts with $\sim 4\cdot
10^{-2}$ ($\sim 10^{-2}$)
at $\tan\beta=2$ and falls down $ 3\cdot 10^{-5}$ ($2\cdot 10^{-4}$) at $\tan\beta\sim 30$. 
These results generally agree  with those of \cite{pilaftsis} though the computational schemes are different.   

Since the parameter space is too wide to cover fully, in the following we restrict ourselves to the following set
\bea
M_{\tilde{L}}=M_{\tilde{R}}=500\,\mbox{GeV}\;, |A_{t}|=1\,\mbox{TeV}\;, |\mu|=250\,\mbox{GeV}\;, \tilde{M}_{A}=200\,\mbox{GeV}
\eea
and vary $\gamma$ over its full range. Each time we consider low and high $\tan\beta$ regimes seperately by taking 
$\tan\beta=4$ and $30$. As $\tan\beta$ increases $|\mu|\cot\beta$ decreases, and this  
enhances the contribution of the radiative corrections. However, variation with $\tan\beta$ is not the whole story
because even for $\cot\beta\leadsto 0$, $\Delta$, $\Delta_{11}$, $\Delta_{12}$ are proportional to $|\mu||A_{t}|$  
so that choice for the latter affects the strength of the radiative corrections. Especially for large $\tan\beta$, stop 
masses weakly depend on $|\mu|$; therefore, these elements of the mass-squared matrix become more
sensitive to the choice for $|\mu|$. The parameter set (29) is a moderate choice in that it enhances the radiative 
corrections through large $|A_{t}|$ term without causing too big splittings among the the $\Delta$ coefficients in (22)-(25)
thanks to the relatively small $|\mu|$ term. Dependence on the parameter $\tilde{M}_{A}$ is as in the CP--invariant theory, 
namely, heavy scalars have masses around $\tilde{M}_{A}$.

\begin{figure}[htb]
\centerline{
\psfig{figure=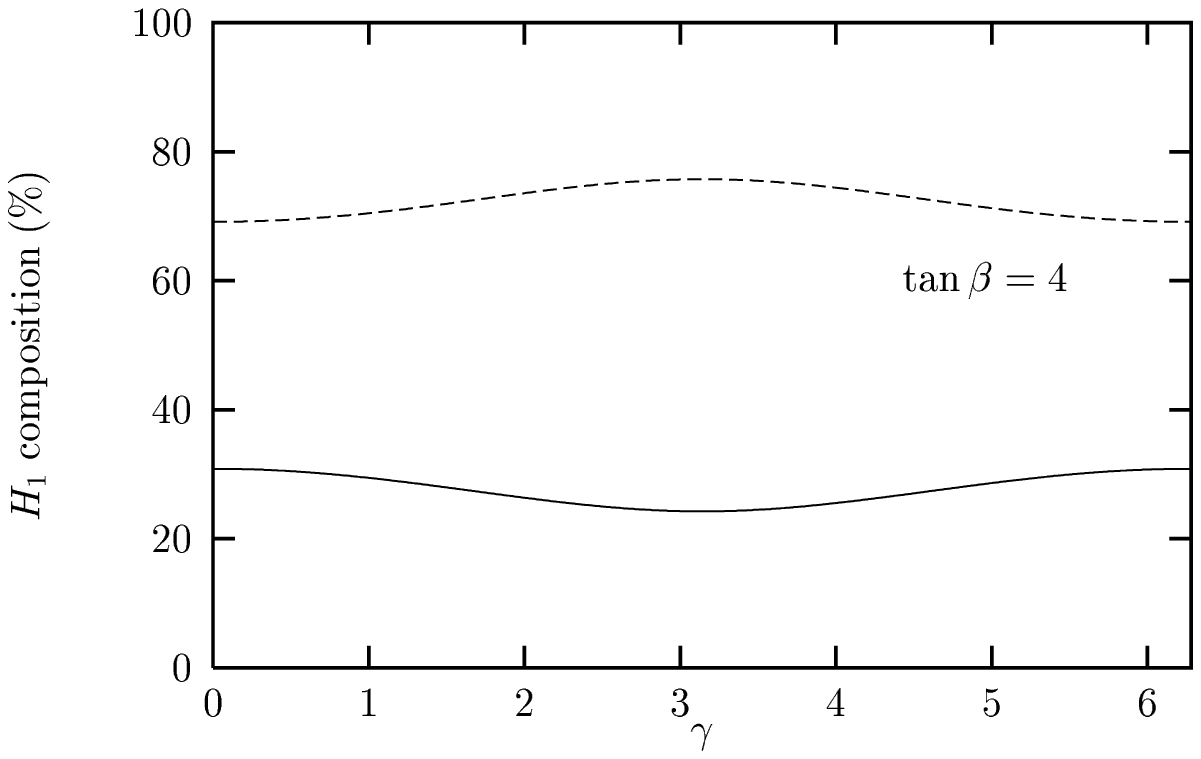,height=10cm,width=8cm,bbllx=-1.cm,bblly=6.cm,bburx=18.cm,bbury=21.cm}
\psfig{figure=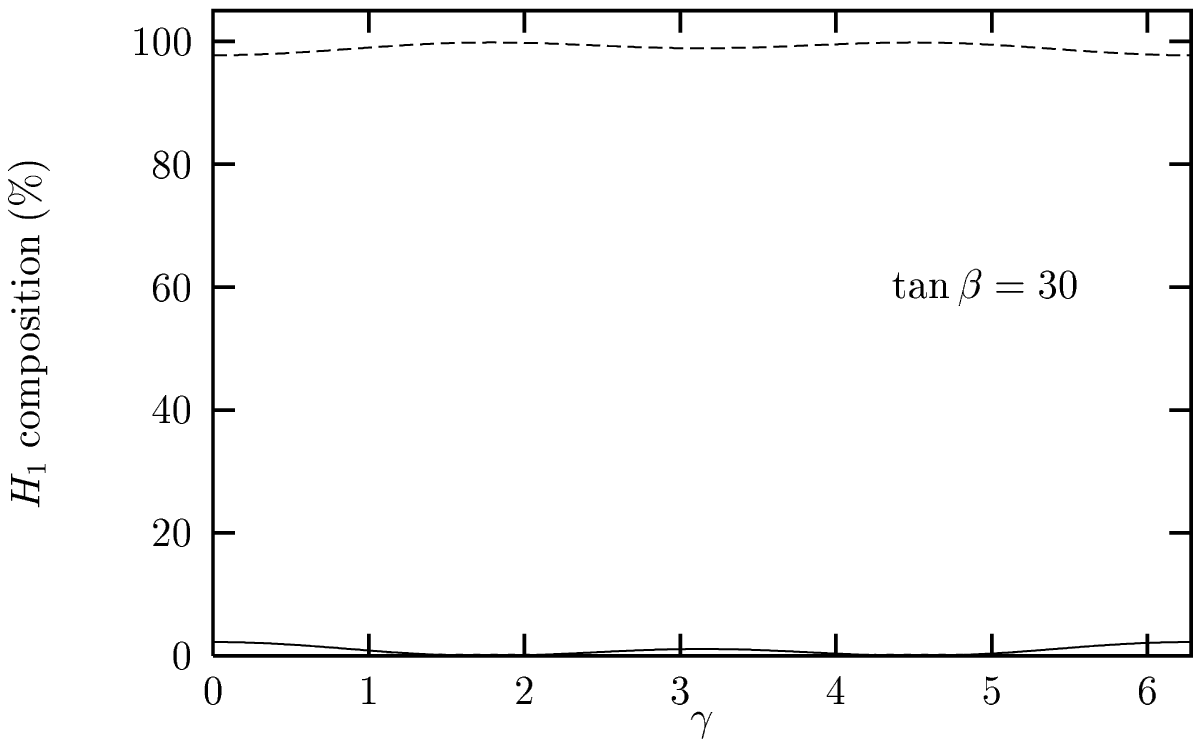,height=10cm,width=8cm,bbllx=1.cm,bblly=6.cm,bburx=20.cm,bbury=21.cm}}
\caption{Percentage composition of $H_{1}$ as a function of $\gamma$ for $\tan\beta=4$ (left panel)
and $\tan\beta=30$ (right panel). Here $\phi_{1}$, $\phi_{2}$ and 
$\sin\beta\varphi_{1}+\cos\beta\varphi_{2}$ contributions are $|{\cal{R}}_{11}|^{2}$ (solid curve), 
$|{\cal{R}}_{12}|^{2}$ (dashed curve) and $|{\cal{R}}_{13}|^{2}$ (short-dashed curve), in percents. 
Values of the parameters are given in (29).}
\end{figure}

\begin{figure}[htb]
\centerline{
\psfig{figure=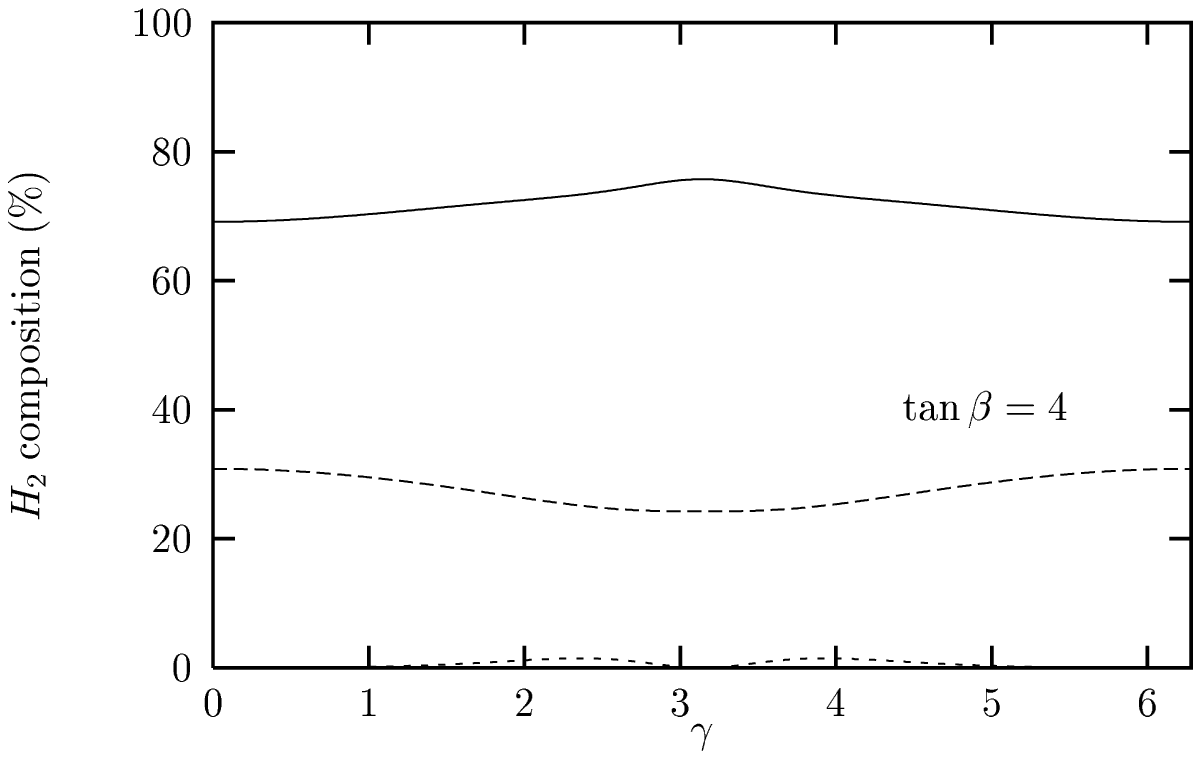,height=10cm,width=8cm,bbllx=-1.cm,bblly=6.cm,bburx=18.cm,bbury=21.cm}
\psfig{figure=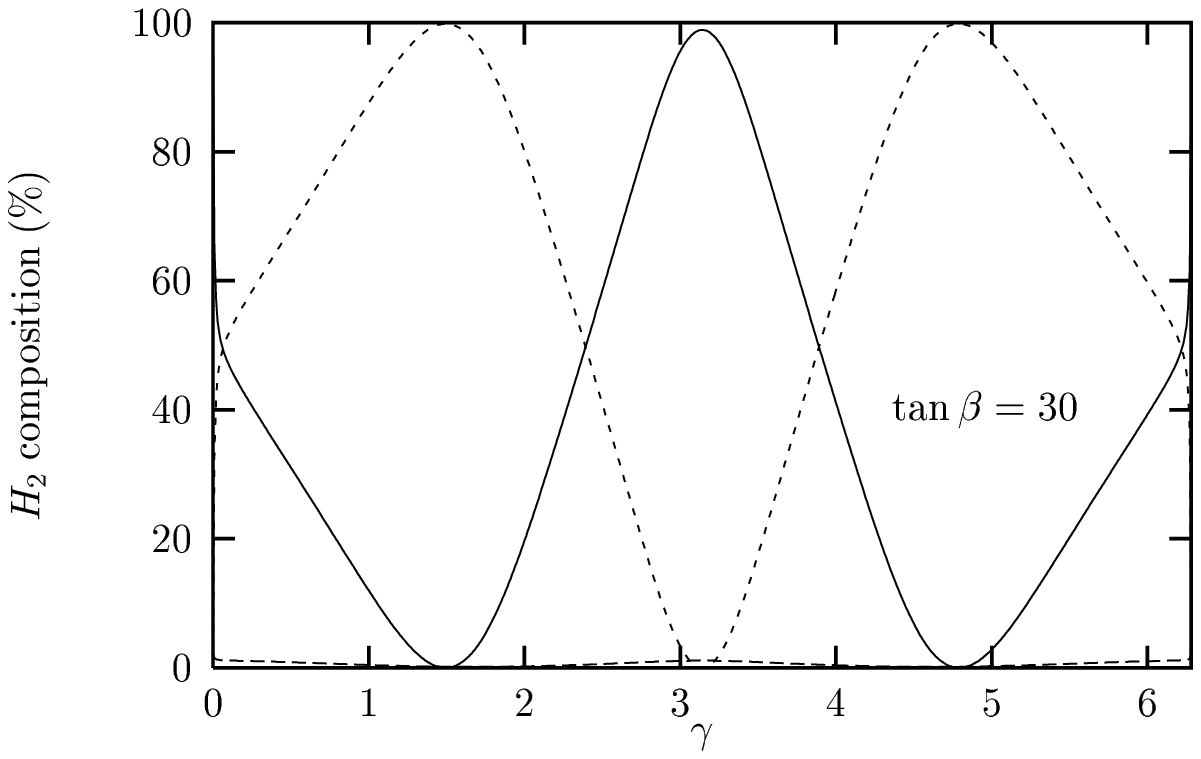,height=10cm,width=8cm,bbllx=1.cm,bblly=6.cm,bburx=20.cm,bbury=21.cm}}
\caption{Percentage composition of $H_{2}$ as a function of $\gamma$ for $\tan\beta=4$ (left panel)
and $\tan\beta=30$ (right panel). Here $\phi_{1}$, $\phi_{2}$ and
$\sin\beta\varphi_{1}+\cos\beta\varphi_{2}$ contributions are $|{\cal{R}}_{21}|^{2}$ (solid curve),
$|{\cal{R}}_{22}|^{2}$ (dashed curve) and $|{\cal{R}}_{23}|^{2}$ (short-dashed curve), in percents.}
\end{figure}

In principle one can diagonalize analytically the scalar mass-squared matrix (19); however, the results
will be too complicated to be suggestive. Instead of using such oblique expressions we shall fix the notation for
diagonalization and numerically analyze the results. The scalar mass-squared matrix (19) can be
diagonalized by a similarity transformation 
\bea
{\cal{R}}\cdot M^{2} \cdot {\cal{R}}^{T}=diag.(M_{H_{1}}^{2}, M_{H_{2}}^{2}, M_{H_{3}}^{2})\;,
\mbox{where}\; 
{\cal{R}} \cdot {\cal{R}}^{T}=1
\eea
where the mass- eigenstate scalar fields are defined by 
\bea 
\left(\begin{array}{c} H_{1}\\ H_{2} \\ H_{3} \end{array}\right)={\cal{R}}\cdot
\left(\begin{array}{c} \phi_{1}\\ \phi_{2}\\ \sin\beta
\varphi_{1}+\cos\beta\varphi_{2} \end{array}\right)
\;.
\eea
$\gamma$-- dependence of the elements of ${\cal{R}}$ is crucial for determining the CP-- impurity of the mass eigenstate 
scalars $H_{i}$. In analyzing ${\cal{R}}$ we adopt a convention such that in the limit of vanishing $\sin \gamma$ we let
$H_{1}\rightarrow h$, $H_{2}\rightarrow H$ and $H_{3}\rightarrow A$, that is, $H_{1}$ is the lightest Higgs. We expect
results of the CP--invariant theory be recovered at the CP--conserving points $\gamma=0,\pi,2\pi$ except for $\gamma$--
dependence of various parameters. 

Depicted in Fig. 2 is the $\gamma$-- dependence of $H_{1}$ composition in percents for the parameter set in eq. (29).
From the left panel we observe that, on the average, $H_{1}$ has $\sim 30\%$ $\phi_{1}$ and $\sim 70\%$ $\phi_{2}$ composition.
As is immediate from the figure, CP=-1 component of $H_{1}$ is small, in fact, it never exceeds $0.02\%$ in the entire range of
$\gamma$. From the right panel, however, we observe that $\phi_{2}$ contribution rises near to  $100\%$ line,
and correspondingly, $\phi_{1}$ contribution remains below $2.5\%$. This result is a consequence of $\beta\rightarrow \pi/2$
limit reminescent from the CP--invariant theory. In this large $\tan\beta$ limit, $\sin\beta \varphi_{1}+\cos\beta\varphi_{2}$
composition of $H_{1}$ reaches at most to $0.2\%$ over the entire range of $\gamma$. It is clear that both windows of the figure
suggest that the lightest Higgs remains essentially a CP--even Higgs scalar for the parameter space in (29).

\begin{figure}[htb]
\centerline{
\psfig{figure=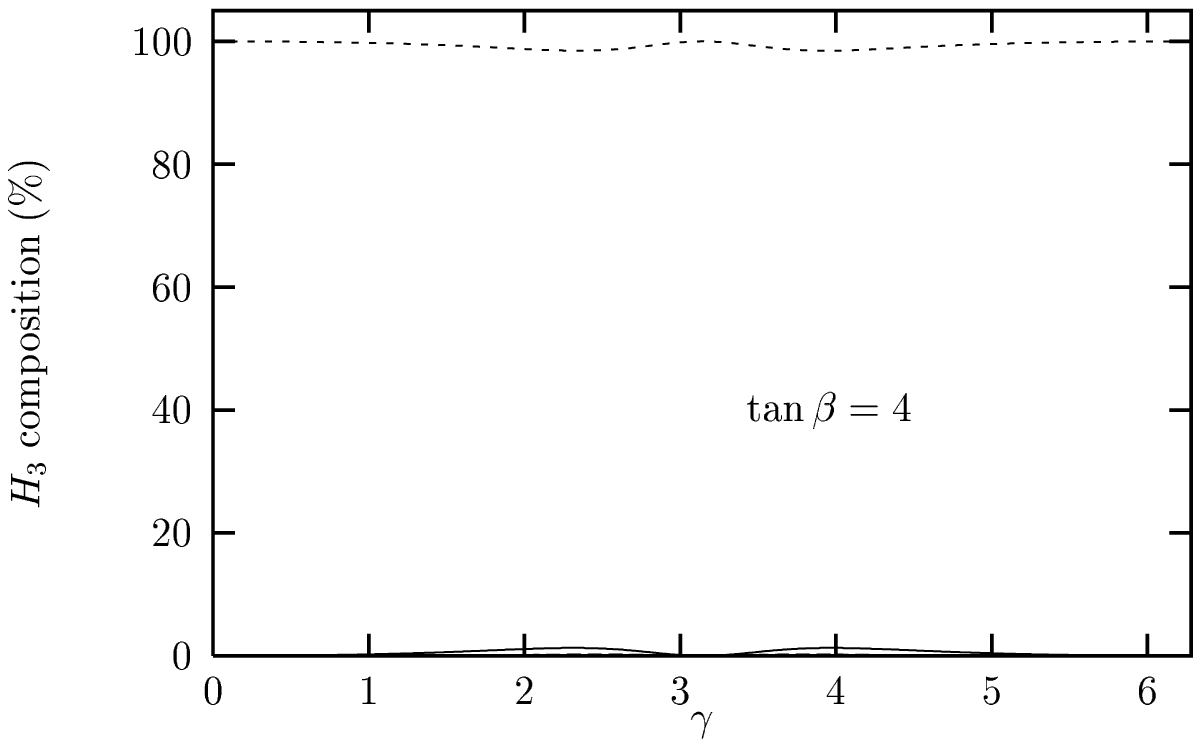,height=10cm,width=8cm,bbllx=-1.cm,bblly=6.cm,bburx=18.cm,bbury=21.cm}
\psfig{figure=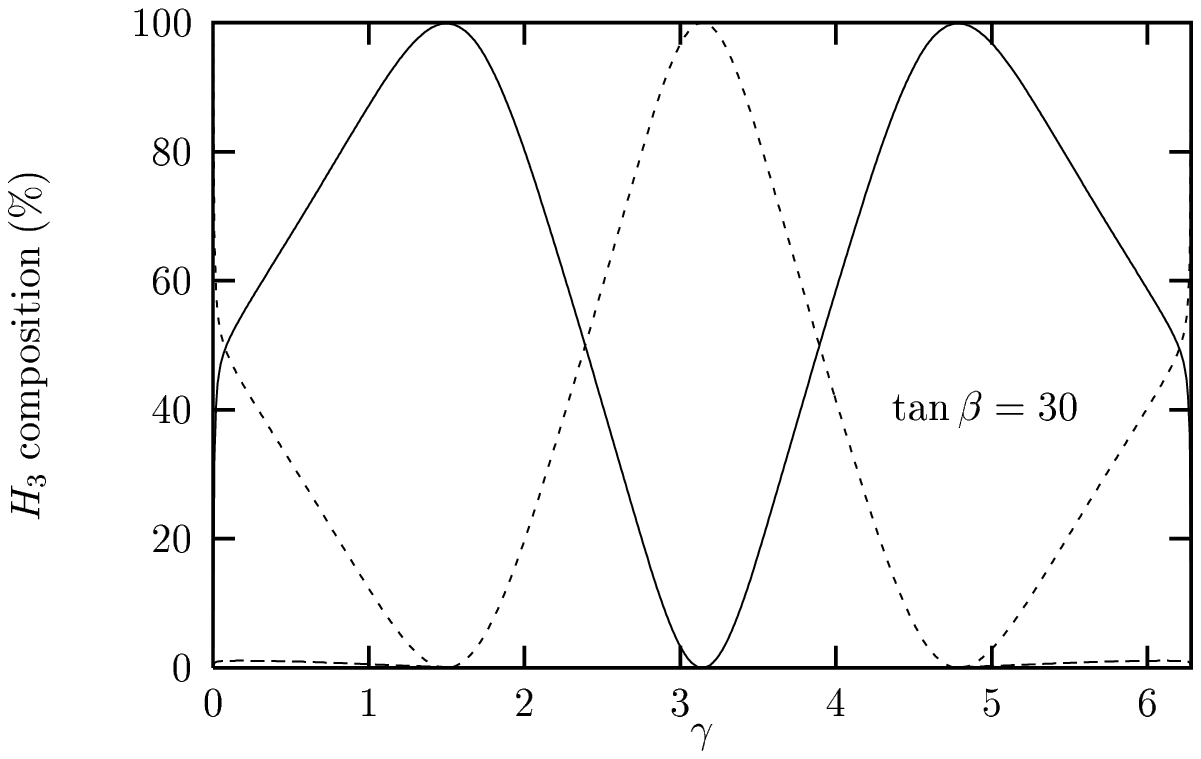,height=10cm,width=8cm,bbllx=1.cm,bblly=6.cm,bburx=20.cm,bbury=21.cm}}
\caption{Percentage composition of $H_{3}$ as a function of $\gamma$ for $\tan\beta=4$ (left panel)
and $\tan\beta=30$ (right panel). Here $\phi_{1}$, $\phi_{2}$ and
$\sin\beta\varphi_{1}+\cos\beta\varphi_{2}$ contributions are $|{\cal{R}}_{31}|^{2}$ (solid curve),
$|{\cal{R}}_{32}|^{2}$ (dashed curve) and $|{\cal{R}}_{33}|^{2}$ (short-dashed curve), in percents.}
\end{figure}

\begin{figure}[htb]
\centerline{
\psfig{figure=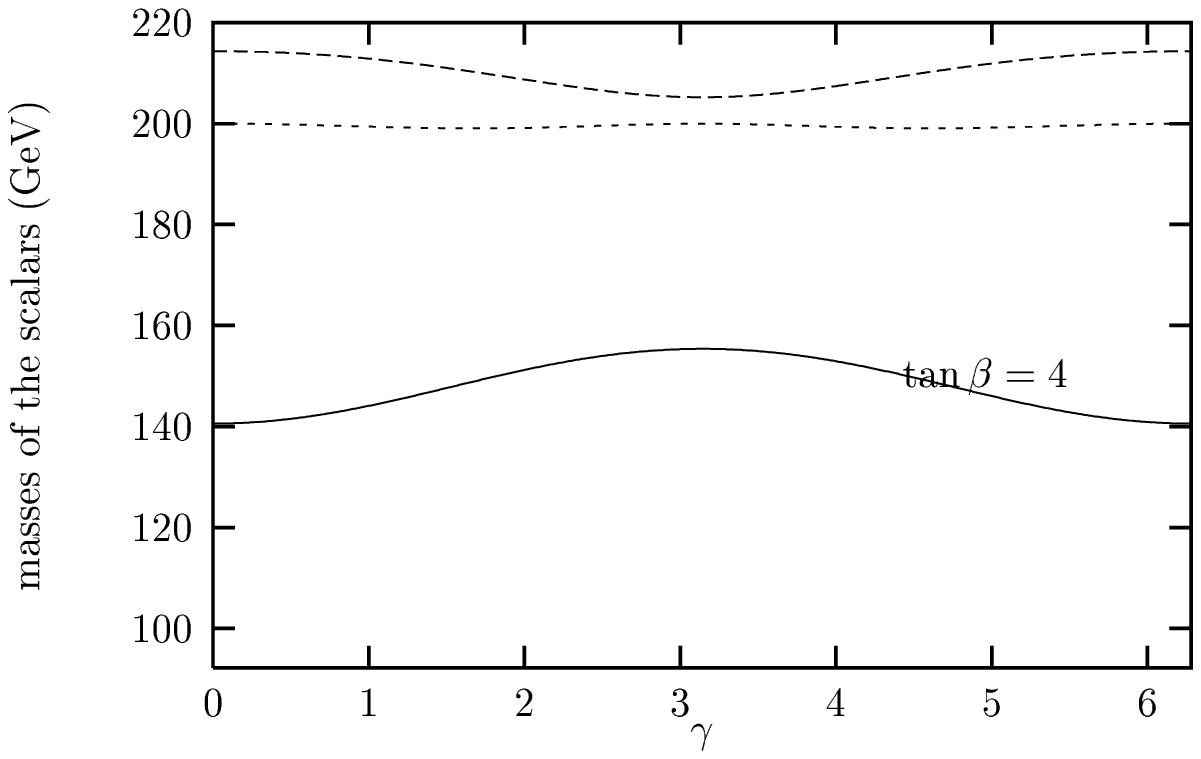,height=10cm,width=8cm,bbllx=-1.cm,bblly=6.cm,bburx=18.cm,bbury=21.cm}
\psfig{figure=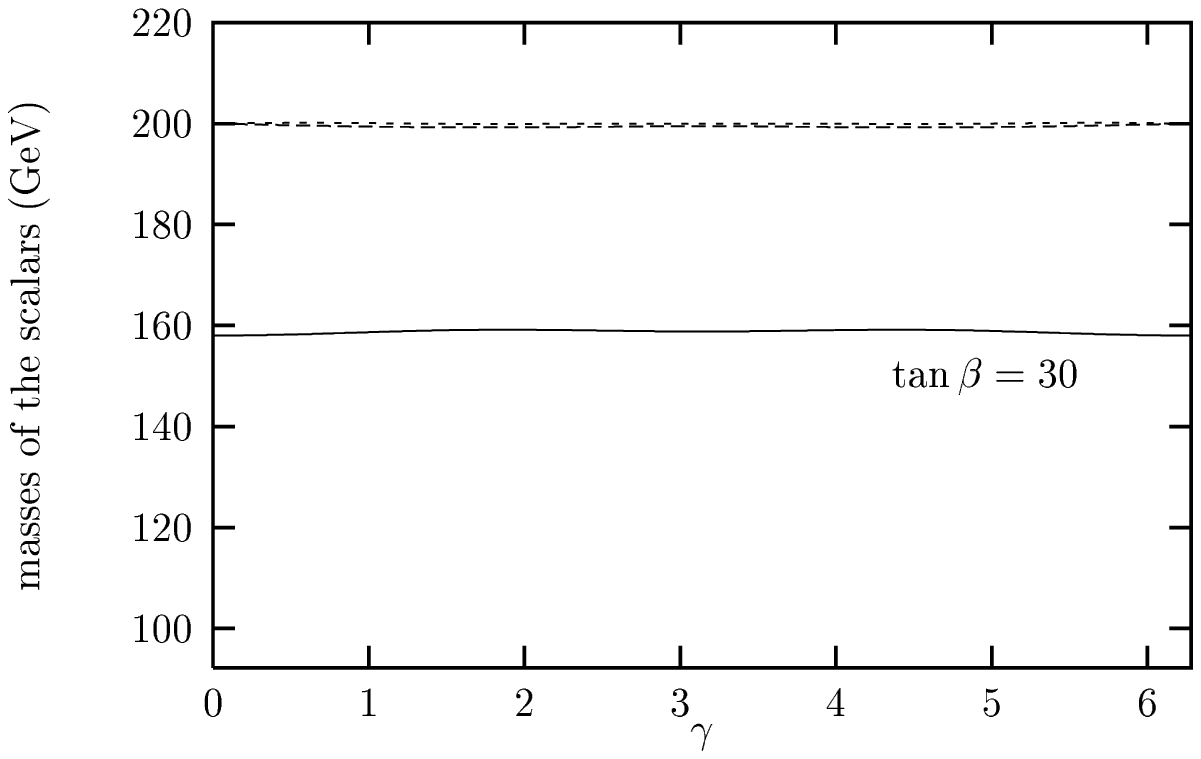,height=10cm,width=8cm,bbllx=1.cm,bblly=6.cm,bburx=20.cm,bbury=21.cm}}
\caption{Masses of the scalars $H_{i}$ as a function of $\gamma$ for $\tan\beta=4$ (left panel) and 
$\tan\beta=30$ (right panel). Here $M_{H_{1}}$, $M_{H_{2}}$ and $M_{H_{3}}$ are shown by solid, 
dashed, and short-dashed curves, respectively. In both panels $H_{1}$ is the lightest scalar whose 
composition is shown in Fig. 2.} 
\end{figure}

In Fig. 3 we show the percentage composition of $H_{2}$ as a function of $\gamma$ for $\tan\beta=4$
(left panel) and $\tan\beta=30$ (right panel). In agreement with the left panel of Fig. 2, for $\tan\beta=4$
(left panel) $H_{2}$ has  $\sim 70\%$ $\phi_{1}$ and $\sim 30\%$ $\phi_{2}$ composition. Unlike $H_{1}$, however,
$\sin\beta \varphi_{1}+\cos\beta\varphi_{2}$ composition of $H_{2}$ becomes as large as $1.3\%$. As expected, 
this increase in the CP=-1 component is compansated by $H_{3}$. More spectacular side of Fig. 3 arises for large 
$\tan\beta$ (right panel) in which $H_{2}$ is seen to gain non-negligible CP--odd composition. In accordance with 
large $\tan\beta$ limit described by eq. (28) there is a strong dependence $\gamma$. Sum of $\phi_{1}$ and $\phi_{2}$
compositions of $H_{2}$ starts from $\gamma=0$ at $100\%$ line, and the former is diminished rather fast 
until $\gamma=\pi/2$. Beyond this point it rises rapidly to $98\%$ line at $\gamma=\pi$ where its CP=+1 component
completes to $100\%$ in accordance with the CP conservation. One notes the complementary behaviour of $\sin\beta
\varphi_{1}+\cos\beta\varphi_{2}$ composition which, in particular, implies that $H_{2}$ is a pure pseudoscalar 
around $\gamma=\pi/2$. 
 
Depicted in Fig. 4 is the percentage composition of $H_{3}$ as a function of $\gamma$ for $\tan\beta=4$ (left panel) and
$\tan\beta=30$ (right panel). In agreement with the left panels of Figs. 2 and 3, $H_{3}$ is almost a pure  pseudoscalar
for $\tan\beta=4$. On the other hand, for $\tan\beta=30$ (right panel) $H_{3}$ is seen to loose its CP--purity in accordance 
with the right panel of Fig. 3. Thus $H_{3}$, except for the  points discussed above, does not have definite CP 
characteristics. From Fig. 3 and Fig. 4 one concludes that heavy scalars have non-negligible CP--impurity in agreemnet 
with the results of \cite{pilaftsis}. 

$\phi_{1}$ (solid curve), $\phi_{2}$ (dashed curve) and $\sin\beta \varphi_{1}+\cos\beta\varphi_{2}$ (short-dashed curve)
compositions of the Higgs fields $H_{i}$ shown in Figs. 2-4   need further elaboration. In accordance with Fig. 1 CP--violating 
mixings are small for small $\tan\beta$ and one necessarily recovers the results of the CP--invariant theory where $\phi_{1}$ and
$\phi_{2}$  mix with each other to form $h^{0}$ and $H^{0}$, and $\sin\beta \varphi_{1}+\cos\beta\varphi_{2}$ is nothing but the 
psedoscalar $A^{0}$. Thus, in this limit mixings are as in the CP--invariant theory as is evident from the left panels 
of each figure. For large $\tan\beta$, however, each Higgs field undergoes certain variations in its compositions. First of all, 
dominant $\phi_{2}$ composition of the light Higgs is easily understandable since the analog of the tree-level Higgs mixing angle (4)
approaches $\beta+\pi/2\approx \pi$ for large $\tan\beta$ \cite{haber}. The remaining component of $H_{1}$ comes mainly form
$\phi_{1}$ since CP--breaking contributions are small for this eigenvalue (See (28)). On the other hand, compositions of $H_{2}$ and
$H_{3}$ are determined from large $\tan\beta$ (dominant $\phi_{1}$ contribution to their CP--even parts) as well as the radiative
corrections. As is evident from (28), in large $\tan\beta$ regime $M_{13}\sim \sin \gamma \cos \gamma$ and
$M_{23}\sim \sin \gamma $ so that (for example) the zeroes of the components of $H_{2}$ follow certain combinations of these 
functional behaviours. The sharp changes in the $\phi_{2}$ and $\sin\beta \varphi_{1}+\cos\beta\varphi_{2}$ compositions of 
$H_{2}$ and $H_{3}$ at $\gamma=0$ (for large $\tan\beta$) follows from the $\gamma$ dependencies of $M_{13}$ and
$M_{23}$ (See eq. (28)). Indeed, variation of the compositions near $\gamma=0$ behaves roughly as $\sin 2\gamma \pm \sin \gamma$ which 
varies quite fast near the origin. 

Until now in Figs. 2-4 we have discussed the CP properties of the eigenstates of the scalar mass-squared matrix (19). Now we analyze
the $\gamma$ dependence of the masses of these scalars for identifying their hierarchy. Fig. 5 shows the $\gamma$--dependence of
the scalar masses for $\tan\beta=4$ (left panel) and $\tan\beta=30$ (right panel). It is clear that $H_{1}$ is the lightest
scalar in both cases. Moreover, the small gap (at most $\sim 20\; GeV$) between $M_{H_{3}}$ and $M_{H_{2}}$ in the left panel is
closed in the right panel where $H_{2}$ and $H_{3}$ are degenerate in mass. This behaviour occurs also in the CP--conserving
case \cite{hunter,radiative1} due to the large value of $\tan\beta$. One more thing about Fig. 5 is that the masses
of $H_{2}$ and $H_{3}$ are almost completely determined by $\tilde{M}_{A}=2\cdot M_{Z}$. For instance, for $\tilde{M}_{A}=10\cdot
M_{Z}$, $H_{2}$ and $H_{3}$ weigh approximately $10\cdot M_{Z}$. Unlike this strong $\tilde{M}_{A}$-- dependence of $M_{H_{3}}$
and $M_{H_{2}}$, mixing among the Higgs scalars and lightest Higgs mass depends mainly on $\tan\beta$. The $\gamma$--dependence 
of the masses around $\gamma=\pi$ (especially in the case of $\tan\beta=4$) differ from those at other CP--conserving points  
due mainly to the minimization of the light stop mass (See eqs. (16) and (17)). Finally, one observes that the light Higgs 
mass is higher than the usual constrained MSSM bounds \cite{jose} due to the large value of $A_{t}$ and relatively small 
soft stop masses which is not possible to produce through RGE's (except, possibly, with non-universal initial conditions).

From Figs. 2-5 one concludes that the lightest Higgs scalar remains essentially a CP=+1 scalar, and  the remaining heavy
scalars do not possess definite CP characteristics. In this sense, MSSM is seen to accomodate a light CP=+1 Higgs boson as in the
CP--invariant case. In addition, the heavy indefinite CP Higgs scalars of the MSSM not only have no correspondent in the SM but also
are distinguishable from the lightest Higgs due to both their masses and CP properties. In the next section we shall discuss decay
properties of these Higgs scalars to fermions and investigate the deviations from the CP--conserving limit. 
 
\section{An example: Decays of the Higgs scalars to fermion pairs}
 
The form of the scalar mass-squared matrix (19) as well as the mixing matrix ${\cal{R}}$ shows
clearly the CP--violation in the Higgs sector. In general, all parameters of the Higgs sector turn
out to depend on these CP--violating angles, for instance, $\tan\beta$, masses of scalars, stop masses are
explicit functions of $\gamma$. In this sense couplings of the Higgs scalars to the MSSM particle spectrum are
necessarily modified. Therefore, the CP--violating angle $\gamma$ can show up, for example, in
the rates for various collision processes testable at future colliders. As an example, one can
consider an $e^{+}e^{-}$ collider with sufficiently large center-of-mass energy. In such a collider
one of the main Higgs production mechanisms is the Bjorken process $Z^{*}\rightarrow Z H_{i}$. When
the center-of-mass energy is varied over a range of values including the masses of the scalars it is
expected that the cross section, as a function of the invariant mass flow into $H_{i}$ branch,
should show three distinct peaks situated at the scalar masses $M_{H_{i}}$. Needless to say, in the CP--
conserving case there would be two peaks instead of three. 
 
We now discuss couplings of the scalars to fermion pairs in detail. That the scalars $H_{i}$
are devoid of definite CP properties influence their couplings to fermions significantly. Rephasing the
fermion fields appropriately one can always make fermion masses real after which coupling of the
scalar $H_{i}$ to $u$-type quarks, $d$-type quarks and charged leptons take the form 
\bea
H_{i}\bar{u}u\; :\; (\sqrt{2} G_{F})^{1/2}\frac{m_{u}}{\sin\beta}({\cal{R}}_{i 2} + i \cos\beta
{\cal{R}}_{i 3} \gamma_{5})\\
H_{i}\bar{d}d\; :\; (\sqrt{2} G_{F})^{1/2}\frac{m_{d}}{\cos\beta}({\cal{R}}_{i 1} + i \sin\beta
{\cal{R}}_{i 3} \gamma_{5})\\
H_{i}\bar{\ell}\ell\; :\; (\sqrt{2} G_{F})^{1/2}\frac{m_{\ell}}{\cos\beta}({\cal{R}}_{i 1} + i
\sin\beta {\cal{R}}_{i 3} \gamma_{5})
\eea
where one observes that each coupling picks up a $\gamma_{5}$ piece showing its CP=-1 content.
Moreover, as the Feynman rules above dictate $\gamma_{5}$ piece is enhanced for large
$\tan\beta$ ($\cot\beta$) for $u$-type quarks ($d$-type quarks and charged leptons). To have a
quantitative understanding of the effects of $\gamma$ on the Higgs decays to fermion pairs it is
convenient to compute the ratio
\bea
R_{i f}&=&\frac{\Gamma(H_{i}\rightarrow \bar{f} f| \gamma \neq 0)} {\Gamma(H_{i}\rightarrow \bar{f}
f|\gamma = 0)}\\&=&\frac{{({\cal{R}}_{i q})}^{2}(1-4m_{f}^{2}/M_{H_{i}}^{2})^{3/2}+
a_{f}^{2}{({\cal{R}}_{i 3})}^{2}(1-4m_{f}^{2}/M_{H_{i}}^{2})^{1/2}}{{({\cal{R}}_{i q
}(0))}^{2}(1-4m_{f}^{2}/M_{H_{i}}^{2}(0))^{3/2}+
a_{f}^{2}{({\cal{R}}_{i 3}(0))}^{2}(1-4m_{f}^{2}/M_{H_{i}}^{2}(0))^{1/2}}\nonumber
\eea
where $q=2 (1)$ and  $a_{f}=\cos\beta (\sin\beta)$ for $u$-type quarks ($d$-type quarks and charged
leptons). The argument "0" of the quantities in the denominator implies the replacement $\gamma =
0$. It is clear that $\Gamma(H_{i}\rightarrow \bar{f}f)$ consists of phase space factors pertinent
to both CP=+1 and CP=-1 cases seperately.  According to the conventions we apply, 
for $i=1, 2$ ($i=3$) only the first (second) term survives in the denominator. The CP--violating 
MSSM phase, $\gamma$ not only functions in creating the additional terms in the Feynman rules (31-33) but 
also affects the couplings and masses themselves.  

\begin{figure}[htb]
\centerline{
\psfig{figure=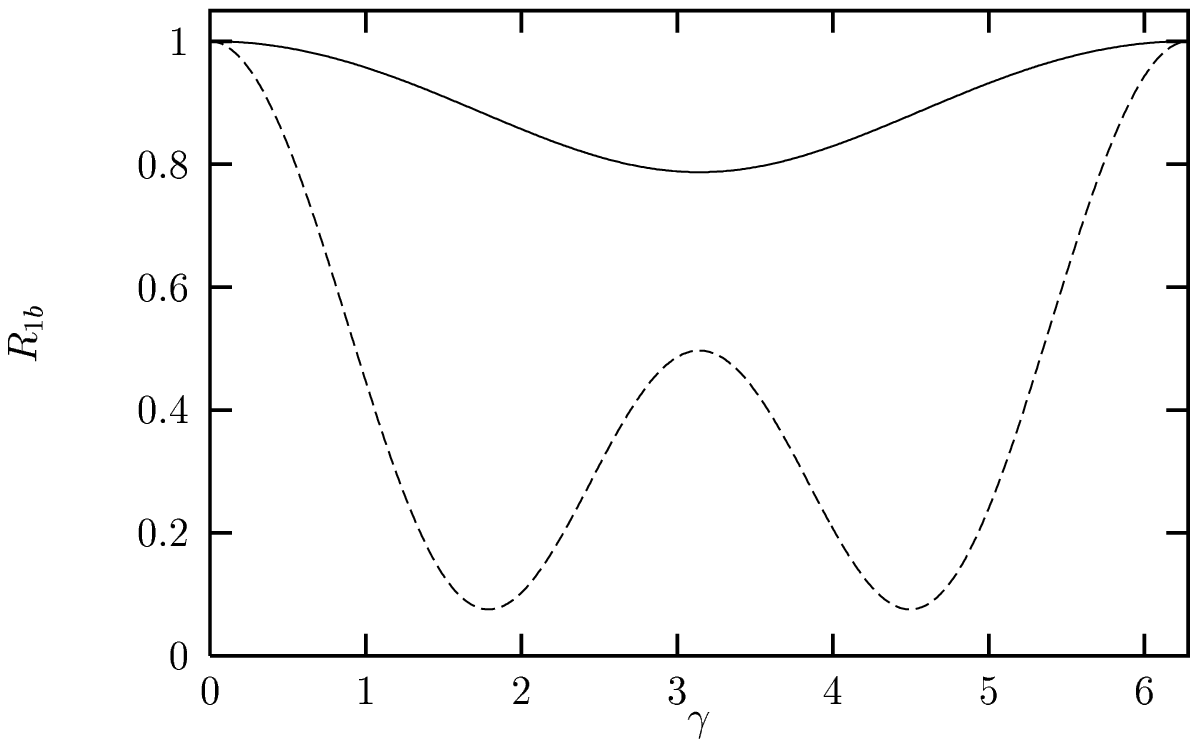,height=10cm,width=8cm,bbllx=-1.cm,bblly=6.cm,bburx=18.cm,bbury=21.cm}
\psfig{figure=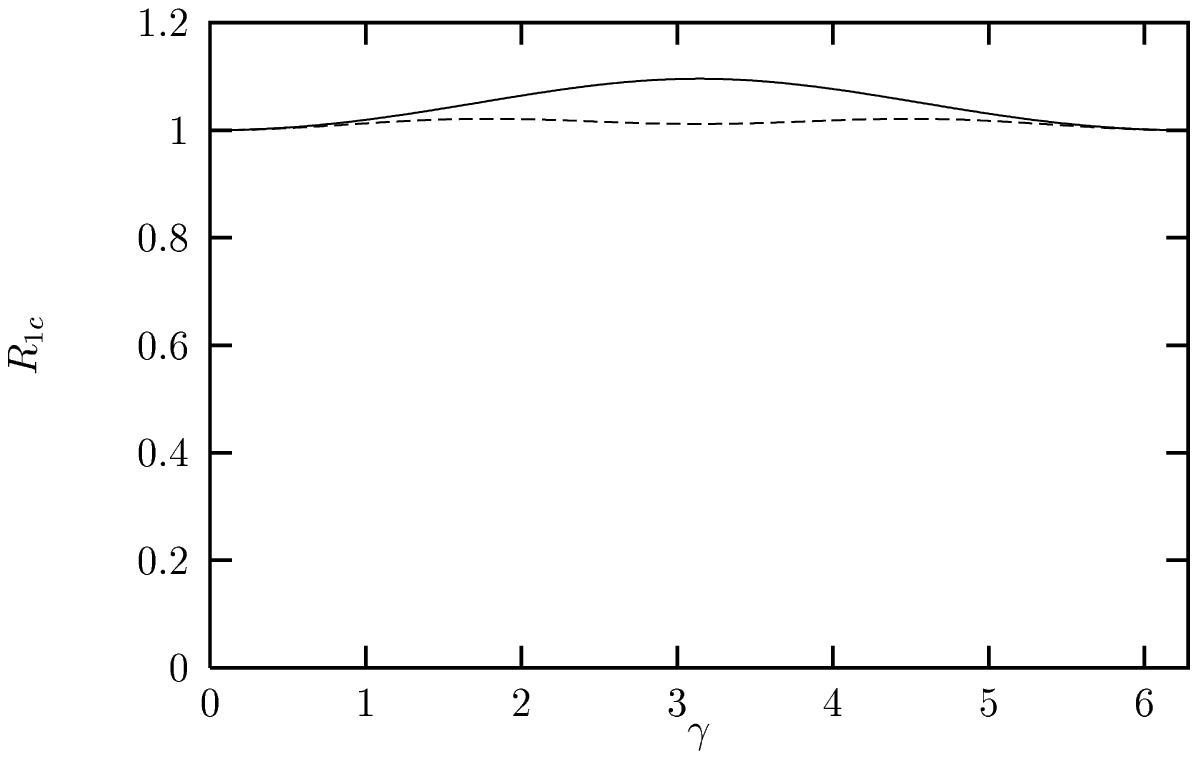,height=10cm,width=8cm,bbllx=1.cm,bblly=6.cm,bburx=20.cm,bbury=21.cm}}
\caption{$R_{1b}$ (left panel) and $R_{1c}$ (right panel) defined in (34) as a function of
$\gamma$ for $\tan\beta=4$ (solid curve) and $\tan\beta=30$ (dashed curve).}
\end{figure}

We now perform a numerical computation of $R_{i f}$ for the parameter space used up to now. In particular, 
we concentrate on two cases $f=b$ and $f=c$, that is, we consider $b$- and $c$- quarks in the computation.
Such an analysis will be exhaustive as it covers the cases listed (32)-(34). We start by listing the quantities
$({\cal{R}}_{ij}(0))^{2}$ necessary for computing $R_{if}$ in Tables 1 and 2. Below, when speaking about the 
elements of the mixing matrices for $\gamma=0$ we shall always refer to these tables.  
\begin{table}[htbp]
\begin{center}
\begin{tabular}{||c|c|c|c||}
$({\cal{R}}_{ij}(0))^{2}$&i=1&i=2&i=3\\ \hline   
$({\cal{R}}_{i1}(0))^{2}$&0.31&0.69&0.0\\ \hline
$({\cal{R}}_{i2}(0))^{2}$&0.69&0.31&0.0\\ \hline
$({\cal{R}}_{i3}(0))^{2}$&0.0&0.0&1.0
\end{tabular}
\end{center}
\caption{\label{table:back1}
Elements of the scalar mixing matrix entering the Feynman rules (32)-(34) for
$\tan\beta=4$ and $\gamma$=0. Here $i$ ($j$) runs over $H_{i}$ ($\phi_{1}, \phi_{2}, \sin\beta\varphi_{1}+\cos\beta\varphi_{2}$).}
\end{table}
\begin{table}[htbp]
\begin{center}
\begin{tabular}{||c|c|c|c||}
$({\cal{R}}_{ij}(0))^{2}$&i=1&i=2&i=3\\ \hline
$({\cal{R}}_{i1}(0))^{2}$&0.02&0.98&0.0\\ \hline           
$({\cal{R}}_{i2}(0))^{2}$&0.98&0.02&0.0\\ \hline
$({\cal{R}}_{i3}(0))^{2}$&0.0&0.0&1.0         
\end{tabular}
\end{center} 
\caption{\label{table:back2}
Elements of the scalar mixing matrix entering the Feynman rules (32)-(34) for 
$\tan\beta=30$ and $\gamma=0$, with the same convention in Table 1.}
\end{table}

\begin{figure}[htb]
\centerline{
\psfig{figure=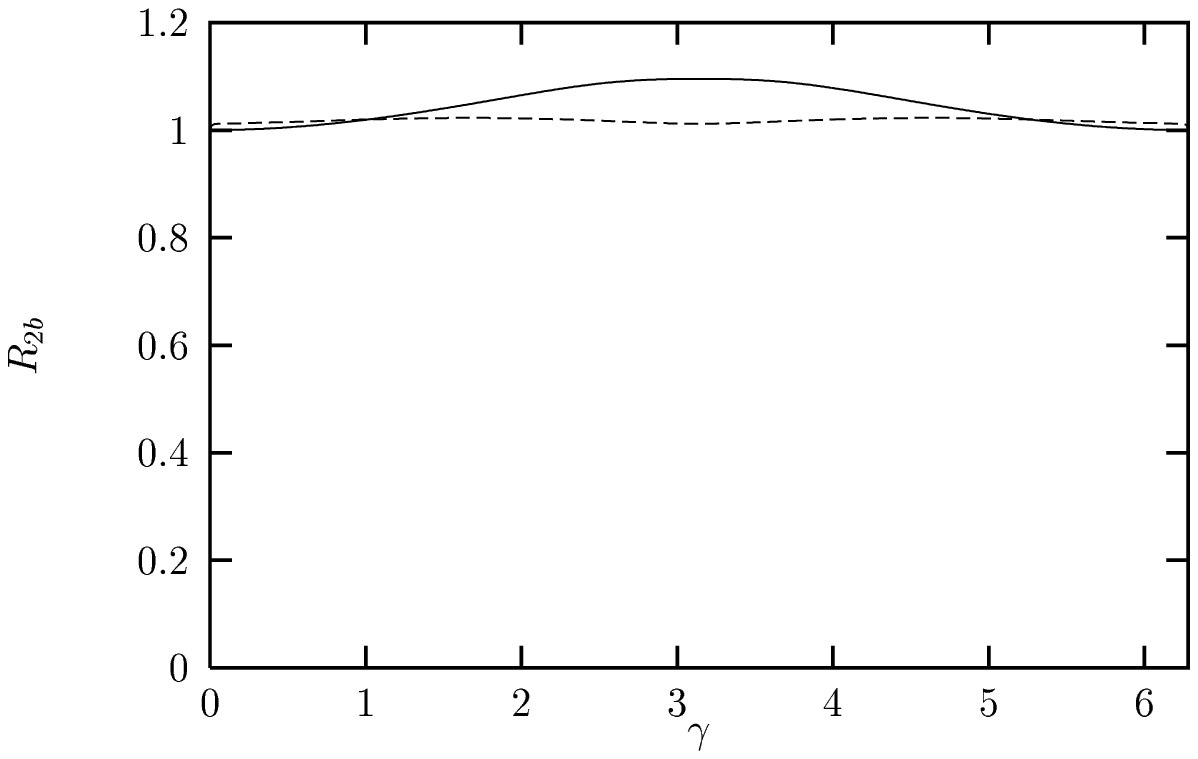,height=10cm,width=8cm,bbllx=-1.cm,bblly=6.cm,bburx=18.cm,bbury=21.cm}
\psfig{figure=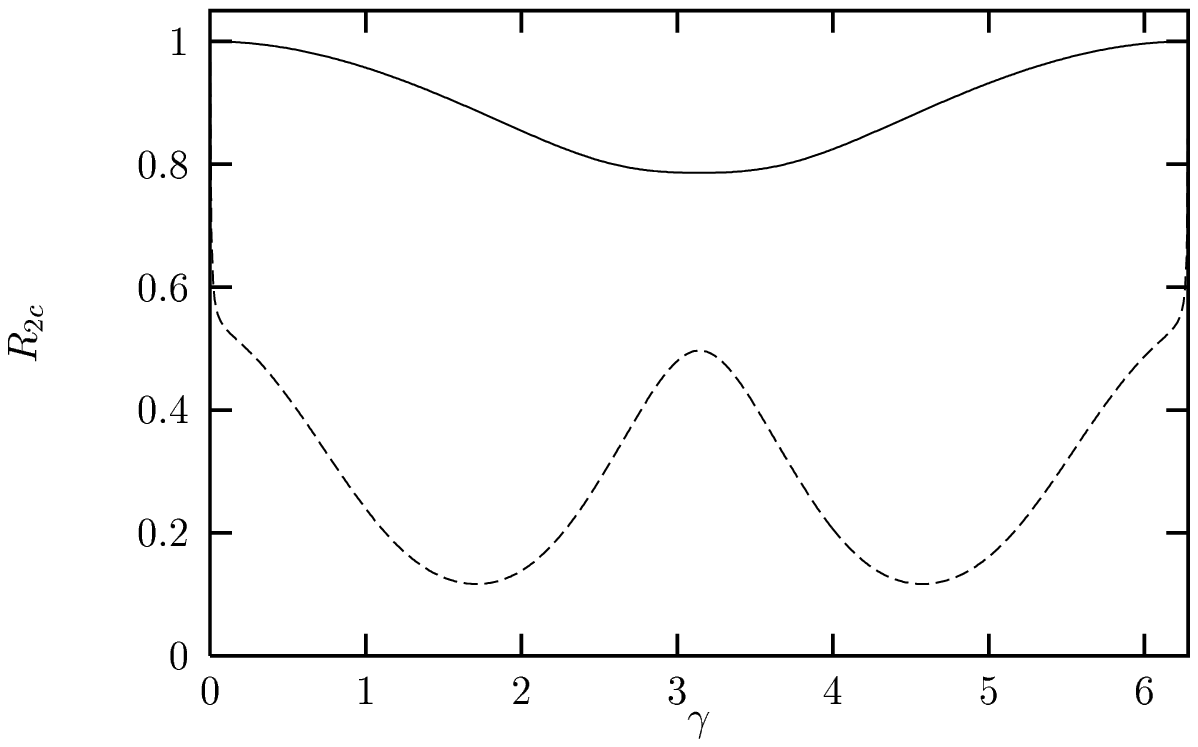,height=10cm,width=8cm,bbllx=1.cm,bblly=6.cm,bburx=20.cm,bbury=21.cm}}
\caption{$R_{2b}$ (left panel) and $R_{2c}$ (right panel) defined in (34) as a function of
$\gamma$ for $\tan\beta=4$ (solid curve) and $\tan\beta=30$ (dashed curve).}
\end{figure}

Fig. 6 shows the $\gamma$--dependencies of $R_{1b}$ and $R_{1c}$ for $\tan\beta=4$ (solid curve) and 
$\tan\beta=30$ (dashed curve). Let us first discuss $R_{1b}$ (left panel). As Fig. 2 shows $\phi_{1}$
composition of $H_{1}$ starts with $0.31$ at $\gamma=0$ (See Table 1) and falls to 0.24 at $\gamma=\pi$.
Therefore, $R_{1b}$ falls to $0.24/0.31= 0.77$ around $\gamma=\pi$. On the other hand, for $\tan\beta=30$ 
$R_{1b}$ obtains a relatively fast variation of $R_{1b}$ with $\gamma$. This behaviour of $R_{1b}$ follows
from its $\phi_{1}$ component in the right panel of Fig. 2, which takes the value of $1.1\%$ around 
$\gamma=\pi$. Therefore, $R_{1b}$ rises to $0.01/0.02=0.5$ at $\gamma=\pi$ as follows from Table 2.
On the other hand, from the right panel of the figure one observes that, $R_{1c}$ remains around 
its conterpart in the CP--conserving limit. Using Tables 1-2, and $\phi_{2}$ compositions in Fig. 2
one can easily infer the behaviour of $R_{1c}$. In particular, constancy of $\tan\beta=30$ curve 
follows from the constancy of the $\phi_{2}$ composition in Fig. 2 (right panel).

\begin{figure}[htb]
\centerline{
\psfig{figure=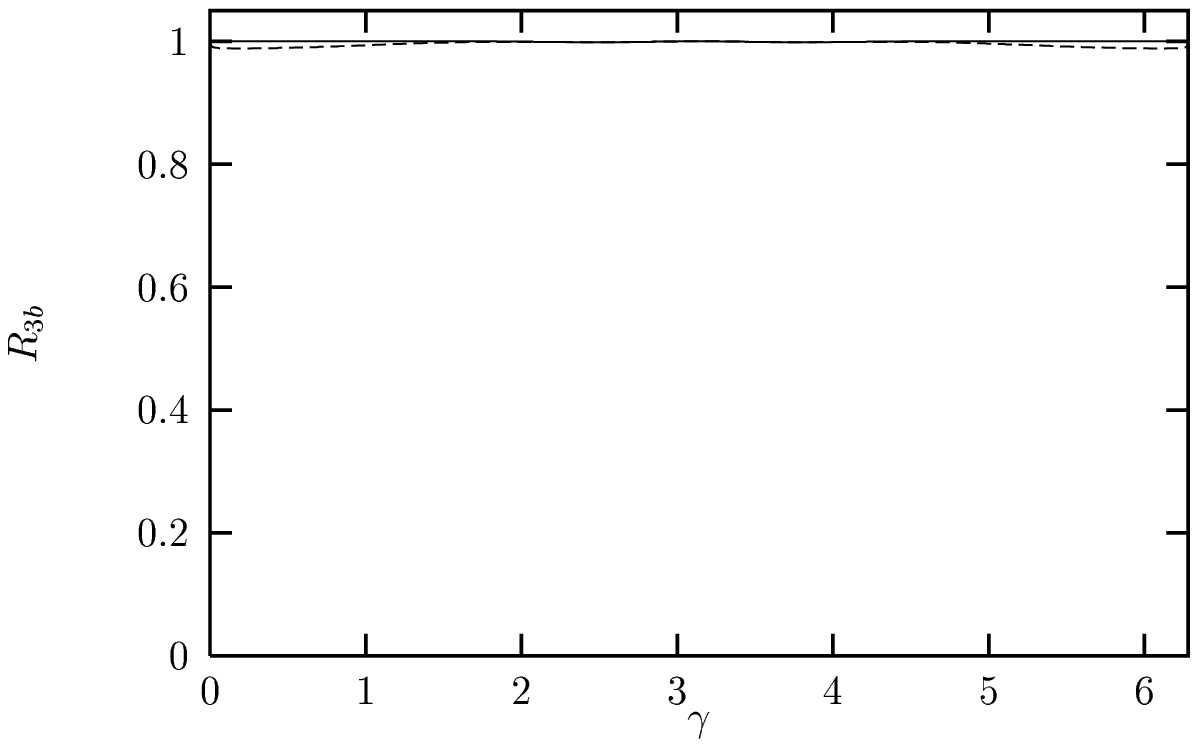,height=10cm,width=8cm,bbllx=-1.cm,bblly=6.cm,bburx=18.cm,bbury=21.cm}
\psfig{figure=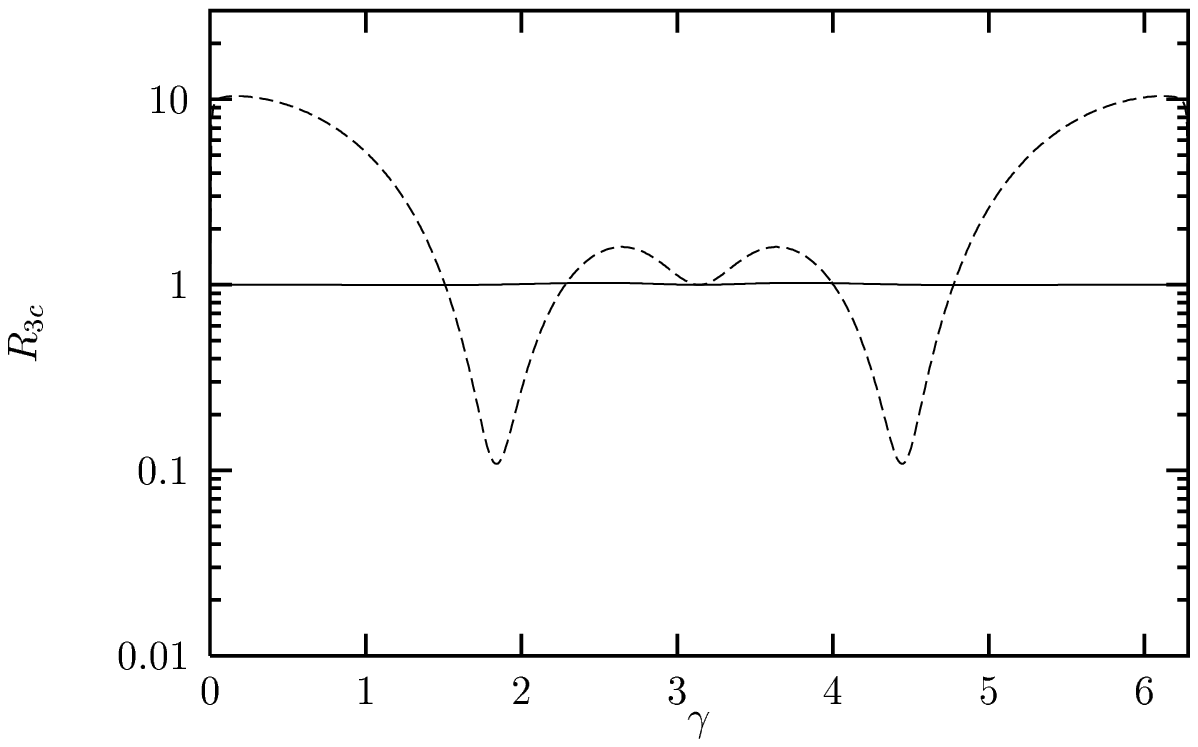,height=10cm,width=8cm,bbllx=1.cm,bblly=6.cm,bburx=20.cm,bbury=21.cm}}
\caption{$R_{3b}$ (left panel) and $R_{3c}$ (right panel) defined in (34) as a function of
$\gamma$ for $\tan\beta=4$ (solid curve) and $\tan\beta=30$ (dashed curve).}
\end{figure}

Depicted in Fig. 7 is the dependence of $R_{2b}$ (left panel) and $R_{2c}$ (right panel)  on $\gamma$ for $\tan\beta=4$ 
(solid curve) and $\tan\beta=30$ (dashed curve). First concentrating on $R_{2b}$, one observes a slow variation with 
$\gamma$ compared to the lightest Higgs (Fig. 6 (left panel)). The behaviour of $\tan\beta=4$ curve simply follows from 
the $\phi_{1}$ composition of $H_{2}$ in Fig. 3 (left panel). However, the flat behaviour of $R_{2b}$ for large $\tan\beta$
follows from the complementary behaviour of its $\phi_{1}$ and $\sin\beta \varphi_{1}+\cos\beta\varphi_{2}$ compositions
shown in Fig. 3 (right panel). It is with the $a_{b}=\sin\beta$ factor in eq. (35) that such a compensation between its
opposite CP--components occur. The variation of $R_{2c}$ follows from the $\phi_{2}$ composition of $H_{2}$ Fig. 3 following 
the same lines of reasoning used in discussing $R_{1b}$ above.

Finally, Fig. 8 shows the variation of $R_{3b}$ (left panel) and $R_{3c}$ (right panel) with $\gamma$ for $\tan\beta=4$
(solid curve) and $\tan\beta=30$ (dashed curve). For $\tan\beta=4$, both $R_{3b}$ and $R_{3c}$ remain around unity because of
the fact that there is little $H-A$ mixing and $({\cal{R}}_{33})^{2}(0)=1$. That $R_{3b}$ remains flat for $\tan\beta=30$
follows from the interplay between its CP=+1 and CP=-1 components as in $R_{2b}$. On the other hand, $R_{3c}$, for $\tan\beta=4$
follows simply from its vanishing $\phi_{2}$ and $100\%$ $\sin\beta \varphi_{1}+\cos\beta\varphi_{2}$ components in Fig. 4 (left
panel). Since $\phi_{2}$ composition is negligibly small for large $\tan\beta$ (right panel of Fig. 4), variation of $R_{3c}$ is 
mainly dictated by its CP=-1 component (short-dotted curve in the right panel of Fig. 4). For example, $R_{2c}=1$ at $\gamma=\pi$
just due to its $100\%$ composition in Fig. 4.  
 
From the study of the decay rates of the Higgs particles to $\bar{b}b$ and $\bar{c}c$ pairs one concludes that their
CP--impurity causes significant changes compared to the CP--invariant limit. In particular, one notes the enhancement in $R_{3c}$
near $\gamma=0$ point, which is one order of magnitude above its value in the CP--conserving limit. 
 
\section{Conclusion and Discussions}
 
In this work we have studied the possible effects of the supersymmetric CP--violating phases on the neutral Higgs scalars of the
MSSM. We have adopted effective potential approximation in computing the radiative corrections and have taken into account 
only the dominant top quark and top squark loops. We now itemize the main results of the work:
\begin{itemize}
\item Radiative corrections induce an unremovable relative phase between the two Higgs doublets. This relative phase remains
non-vanishing as long as the supersymmetric CP--violating phases are finite, and determines the dynamics of the electroweak phase
transition \cite{phasetrans}.
\item The CP=+1 and CP=-1 components of the Higgs doublets are mixed up due to the mixing terms in the scalar mass-squared matrix
which
\begin{enumerate}
\item are proportional to the relative phase between the two doublets,
\item increase with increasing $\tan\beta$. 
\end{enumerate}
\item The lightest Higgs scalar remains essentially CP--even irrespective of the supersymmetric CP phases. Therefore,  
the MSSM has a light CP=+1 scalar as in the CP--respecting case which can, however, be distinguished from that of 
the CP--invariant theory, for example, by its reduced decay rate to $\bar{b}b$ pairs.
 
\item As in the CP--invariant case, there are two heavy scalars which have 
\begin{enumerate}
\item definite CP quantum numbers for small $\tan\beta$ values,
\item no definite CP characteristics for large $\tan\beta$.
\end{enumerate}
\item The strong mixing between the heavy scalars affect significantly their decay rates to fermion pairs, and such mixings can be
important in other collision processes testable at future colliders. Especially the decay rate of the would-be CP--odd scalar gets 
enhanced for $0<\gamma\leq \pi/2$.
\item The supersymmetric CP--violating phases not only cause the creation of CP--violating mixings but also modify the couplings
and masses compared to ones in the CP--invariant case.
\end{itemize}
In the light of these results one concludes that the supersymmetric phases can be useful tools for obtaining 
manifestations of supersymmetry through their effects on collision and decay processes testable in near future colliders.

\section{Acknowledgements}
It is a pleasure for author to express his gratitude to A. Masiero for highly useful discussions and his careful reading of
the manuscript. He would like to thank to T. M. Aliev, too,  for helpful discussions.

\end{document}